\documentclass{article}
\newcommand{\bfr}{\begin{flushright}}
\newcommand{\efr}{\end{flushright}}
 
\begin{document}
\title{Multicentered solution for maximally charged dilaton black 
holes in arbitrary dimensions
}
\author{Kiyoshi Shiraishi\\
Akita Junior College, Shimokitade-Sakura, Akita-shi, Akita 010, Japan
}
\date{J. Math. Phys. 34, 1480--1486 (1993)
}
\maketitle
\begin{abstract}
A family of static multicentered solutions to modified
Einstein-Maxwell equations coupled with a dilaton is constructed in
$(1+N)$ dimensional space-time ($N\ge 2$). For $N\ge 3$, the solutions
are generalizations of the Majumdar-Papapetrou solution. We also find
the solution in $(1+2)$ dimensions, where the scalar and vector forces
cancel each other in the static case. The interaction between two
extreme charged dilaton black holes in the low-energy limit is
investigated in $(1+N)$ dimensions ($N\ge 3$). We find that there
remains the residual velocity-dependent force in general cases, except
for the case with $N=a^2$.
\end{abstract}

\section{INTRODUCTION}
There has recently been a revived interest in the exact solutions in
the coupled system with a dilaton.\cite{1,2,3,4} Although the study of
dilaton-coupled Einstein equations has been motivated by string theory
and Kaluza-Klein theory, it has recently turned out that black holes
described by the solution for the generalized dilaton coupling have
curious properties.\cite{1,2,3,4,5} 

The multi-black-hole solution in
usual Einstein-Maxwell system is known as the Papapetrou-Majumdar
metric,\cite{6} which represents the static equilibrium among extreme
Reissner-Nordstrom black holes. Recently, the multicentered solution in
string theory has been found in Ref.~\cite{2}. 
The difference between
these two solutions originates from the coupling to the dilaton in the
case of string theory. Actually it has been suggested that the extreme-
black-hole solution in string theory has peculiar nature.\cite{2} 

We thus come
to take an interest in interpolating the two cases: In this paper, we
consider the generalized dilaton coupling in modified Einstein-Maxwell
system. We construct the multi-black-hole solutions in various
dimensions. The generalized Papapetrou-Majumdar solutions in $(1+N)$
dimensional Einstein-Maxwell system without the dilaton have been
discussed by Myers.\cite{7} The solutions we obtain in Sec.~2 contains
Myers' solution as a solution in the limit of dilaton decoupling.

Further, we show the static multicentered solution in $(1+2)$
dimensions in Sec.~3. 

In Sec.~4, we study the interaction between
two maximally charged dilaton black holes in the low-energy limit. The
last section is devoted to conclusion. 

\section{THE MULTIDILATON BLACK-HOLE SOLUTION IN FOUR
AND MORE THAN FOUR DIMENSIONS}
 We start with the action
\begin{equation}
S=\int d^{N+1}x\frac{\sqrt{-g}}{16\pi}\left[
R-\frac{4(\nabla\phi)^2}{N-1}-e^{-\{4a/(N-1)\}\phi}F^2\right]\,,
\label{2.1}
\end{equation}
where we set the Newton constant equals to
one. The constant $a$ (which can be taken as non-negative) is the
parameter which determines the strength of the coupling between the
Maxwell field $F$ and the dilaton field $\phi$. For $a=0$, the action
describes the usual Einstein-Maxwell system with a free scalar field: a
charged nonrotating black hole in the system is represented by the
Reissner-Nordstrom solution. For $a=1$, the action reduces to the one
which derived from the low-energy string theory. 

For $N=3$, $a=0$ case,
multicentered static solutions to the Einstein-Maxwell equations were
found by Papapetrou and Majumdar.\cite{6} Further, generalization to
higher dimensions was attained by Myers.\cite{7} Their solutions
correspond to a many-body system of extreme charged black holes, where
the gravitational attraction and the Coulomb repulsion is exactly
canceled. In the present paper, we manage to extend their studies to
the system including the dilaton field. As a result, the scalar force
between black holes come to take part in the system. 

In this section,
we treat the case with $N\ge 3$. We handle the $N=2$ case separately in
the later section, because the solution in $(1+2)$ dimensions behaves
very differently from the ones in higher dimensions. 

The extremity
condition for charged dilaton black holes in arbitrary dimensions can
be read from the results in Refs.~\cite{1}. If we fix the asymptotic
value of the dilaton at spatial infinity to be zero, we flnd that the
extremity condition is expressed as the following ratio of charge to
mass: 
\begin{equation}
\frac{|Q|}{m}=\frac{8\pi}{A_{N-1}}\left(\frac{N-2+a^2}{2(N-1)}
\right)^{1/2}\,,
\label{2.2}
\end{equation}
where $A_{N-1}=2\pi^{N/2}/\Gamma(N/2)$. Please note that the expression
would take a different form in accordance with the choice of the
normalization of the charge. The present notation coincides with
Ref.~\cite{2} when $N=3$ and $a=1$. The extremity condition must be
realized for each point ``source'' in the multi-black-hole solution.
The extremity and balance condition will be con- sidered further in
Sec.~4. 

Now we turn to the multicentered solution. The general metric
ansatz (for $N\ge 3$) can be written in the isotropic
coordinates:\cite{7} 
\begin{equation}
ds^2=-U^{-2}({\bf x}^k)dt^2+U^{2/(N-2)}({\bf x}^k)\delta_{ij}d{\bf x}^i
d{\bf x}^j\,.
\label{2.3} 
\end{equation}
We also assume that the
electric potential and the dilaton configuration are proportional to
some powers of $U({\bf x}^k)$. Solving the equations of motion obtained
from the action (\ref{2.1}) we get the general form of $U({\bf x}^k)$
for an ``$n$-body'' system: 
\begin{equation}
U({\bf x}^k)=\{F({\bf x}^k)\}^{(N-2)/(N-2+a^2)}\,,
\label{2.4}
\end{equation}
where 
\begin{equation}
F({\bf x}^k)=1+\frac{1}{N-2}\sum_{i=1}^n\frac{\mu_i}{|{\bf x}-{\bf
x}^i|^{N-2}}\,,
\label{2.5}
\end{equation}
and using this expression, we
obtain the potential and the dilaton configuration: 
\begin{equation}
A=\pm\left(
\frac{N-1}{2(N-2+a^2)}
\right)^{1/2}\{F({\bf x}^k)\}^{-1}dt
\label{2.6}
\end{equation}
and
\begin{equation}
e^{-\{4a/(N-1)\}\phi}=\{F({\bf x}^k)\}^{2a^2/(N-2+a^2)}\,.
\label{2.7}
\end{equation}
The constant $\mu_i$ has the following connection to the mass and the
charge of each source:
\begin{equation}
m_i=\frac{A_{N-1}(N-1)}{8\pi(N-2+a^2)}\mu_i\,,
\label{2.8}
\end{equation}
\begin{equation}
|Q_i|=\left(
\frac{N-1}{2(N-2+a^2)}
\right)^{1/2}\mu_i\,.
\label{2.9}
\end{equation}
In this solution, we have fixed the asymptotic value for $\phi$ to be
zero. 

The solution obtained here coincides with the solution of
Myers\cite{7} in the $(1+N)$-dimensional Einstein-Maxwell system when
$a=0$; therefore, of course, our solution includes the
Papapetrou-Majumdar solution\cite{6} as a special case for N= 3. The
multi-black-hole solution shown by the authors of Ref.~\cite{2} can be
rederived if $N=3$ and
$a=1$ are substituted in the present solution. 

The discussion on the
physical interpretation of the solution is left for the subject of
Sec.~4.

\section{CHARGED DILATON SOLITON IN $(1+2)$-DIMENSIONAL SPACE-TIME}
In three-dimensional space-time, there is no propagating graviton mode
in Einstein gravity; nevertheless, the global geometry of space-time
is governed by Einstein equations. Various aspects of three-dimensional
Einstein gravity have been investigated by many authors in the past
decade\cite{8,9} (and see also Ref.~\cite{10}). 

In three dimensions,
there is no black-hole solution; thus the extreme charged object in
$(1+2)$ dimension seems to make no sense. The static, charged many-body
system cannot exist because there is no attractive gravitational force
to balance with the Coulomb force. 

There is another story in the
Einstein-Maxwell dilaton system. The scalar force induced by the
dilaton can cancel the electric force. Another prospect is obtained
from the extremity condition. If the condition (\ref{2.2}) can be
extrapolated to the $N=2$ case, it shows $|Q|/m = 2^{3/2}a$. Thus in
the absence of the dilaton, maximal charge of a ``charged source'' is
found to be zero. This interpretation sounds reasonable; the metric for
the static configuration of an arbitrary number of point masses are
known in the three-dimensional Einstein gravity.\cite{8} This
corresponds to the three-dimensional extension of the
Papapetrou-Majumdar solution. 

Now we first get the ``spherical''
solution in the three-dimensional Einstein-Maxwell dilaton system.
The action is given by (\ref{2.1}) with $N=2$ substituted. Using the
isotropic coordi- nates, one can find the exact solution 
\begin{equation}
ds^2=-dt^2+|{\bf x}^2|^{-4m}|f({\bf x})|^{2/a^2}\delta_{ij}d{\bf x}^i
d{\bf x}^j\,,
\label{3.1}
\end{equation}
with
\begin{equation}
e^{-2a\phi}=f({\bf x})=-C\ln(|{\bf x}|/r_c)\,,
\label{3.2}
\end{equation}
\begin{equation}
e^{-4a\phi}F=\frac{Q}{|{\bf x}|}\frac{{\bf x}^i}{|{\bf x}|}dt\wedge
d{\bf x}^i\,,
\label{3.3}
\end{equation}
where $m$, $C$, and $r_c$
are constants. Note that the solution has singularities at a finite
distance $r$, besides the origin. The present solution has no smooth
connection to the $a=0$ case. In Ref.~\cite{9}, the Einstein-Maxwell
equations is solved in three dimensions. Their solution also has the
singularity at a finite distance. 

More notable remark is on $m$, which
can take any value in general. The total mass of the object described
by the solution is not well defined because of the divergence of the
energy of the electric field. Thus we cannot call $m$ as mass in a
precise sense. At the same time, in three dimensions, the asymptotic
flat condition cannot be attained in a practical sense. This is crucial
for studying the properties of any solution to the three-dimensional
Einstein equation. 

The multicentered solution can also be obtained in
the isotropic coordinates. We find that the solution is given by
\begin{equation}
ds^2=-dt^2+G({\bf x})|f({\bf x})|^{2/a^2}\delta_{ij}d{\bf x}^i
d{\bf x}^j\,,
\label{3.4}
\end{equation}
with
\begin{equation}
e^{-2a\phi}=f({\bf x})=-2^{1/2}a\sum_{i=1}^nQ_i\ln(|{\bf x}-{\bf
x}_i|/r_{ci})\,,
\label{3.5}
\end{equation}
\begin{equation}
e^{-4a\phi}F=\sum_{i=1}^n\frac{Q}{|{\bf x}|}\frac{{\bf x}^i}{|{\bf
x}|}dt\wedge d{\bf x}^i\,,
\label{3.6}
\end{equation}
and
\begin{equation}
G({\bf x})=\prod_{i=1}^n(|{\bf x}-{\bf x}_i|/b_i)^{-8m_i}\,. 
\label{3.7}
\end{equation}
Here $m_i$ (and $b_i$) can take arbitrary values. Actually, $\ln G$ is
nothing but the solution to the two-dimensional Laplace equation. The
ambiguous situation come from the fact that we cannot constrain the
behavior of the metric at asymptotic region of space in $(1+2)$
dimensions. 

The naive extension of the extremity condition turns out
wrong. We can, however, find an interesting similarity to the solution
for the other dimensions. To carry out the analysis, we must set $G(x)
=1$ in the solution (\ref{3.4}). This could be fulfilled by imposing an
appropriate condition near the point charge, instead of the asymptotic
behavior. 

The idea is to look into the vicinity where $f\sim1$ around a
point source. In that place, the metric (\ref{3.4}) of space behave as
\begin{equation}
g_{ij}\sim\{1-(2^{3/2}/a)Q_i\ln(|{\bf x}-{\bf
x}_i|/r'_c)\}\delta_{ij}\,,
\label{3.8}
\end{equation}
where $r'_c$ is a constant. On the other hand, we consider a point mass
mi. The space component of the metric is approximately given in the
vicinity of unity: 
\begin{equation}
g_{ij}(\mbox{point mass})=(|{\bf x}-{\bf x}_i|/b_i)^{-8m_i}\delta_{ij}
\sim\{1-8m_i\ln(|{\bf x}-{\bf x}_i|/b_i)\}\delta_{ij}
\label{3.9}
\end{equation}
If we try to identify the behaviors, we must take
\begin{equation}
Q_i/m_i=2^{3/2}a\,.
\label{3.10}
\end{equation}
This is no more than the ``naively'' extended extremity condition! In
higher dimensions, the asymptotic region corresponds to the space
component of the metric $g_{ij}\sim\delta_{ij}$. 

The further
implication of the solution will be exhibited in Sec.~4.

\section{THE INTERACTION BETWEEN TWO MAXIMALLY CHARGED DILATON BLACK
HOLES IN ARBITRARY DIMENSIONS}
First we examine the balance condition in a physical perspective. The
system we have treated is governed by three kinds of forces; the
Newtonian attraction, the Coulomb repulsion, and the attractive scalar
force mediated by the dilaton. We can define the scalar charge $\sigma$
by the asymptotic behavior of the dilaton as
\begin{equation}
\frac{2^{1/2}}{(N-1)^{1/2}}\nabla_i\phi\sim-\frac{\sigma}{|{\bf
x}|^{N-1}}\frac{{\bf x}^i}{|{\bf x}|}\quad\mbox{as}\quad |{\bf
x}|\rightarrow\infty\,,
\label{4.1}
\end{equation}
because of the normalization required from the realization of the
``proper'' kinetic term  (note, however, the normalization in
Ref.~\cite{1} is different from ours; this is due to the difference in
the normalization of the kinetic term for the vector field). The
normalization here is chosen for the correspondence to the Coulomb
potential. 

The Newtonian force is given by
\begin{equation}
(F_N)_i=-m\nabla_i\phi_N\,,
\label{4.2}
\end{equation}
where the Newton potential $\phi_N$ is defined via
\begin{equation}
g_{00}\sim-(1+2\phi_N)\quad\mbox{as}\quad |{\bf
x}|\rightarrow\infty\,,
\label{4.3}
\end{equation}
as the leading term.

Consequently, the amount of the total force between two point sources
$a$ and $b$ separated in the distance is written as
\begin{equation}
\frac{A_{N-1}}{4\pi}\left[\frac{Q_aQ_b}{R_{ab}^{N-1}}-
\frac{\sigma_a\sigma_b}{R_{ab}^{N-1}}-\left(\frac{4\pi}{A_{N-1}}
\right)^2\frac{2(N-2)}{N-1}
\frac{m_am_b}{R_{ab}^{N-1}}
\right]\,,
\label{4.4}
\end{equation}
where $R_{ab}$ is the distance between the point sources $a$ and $b$.
The unpleasant factor $A_{N-1}/4\pi$ appears in our convention, since we
must sweep the dependence on the space-time dimension into somewhere.
One can immediately confirm the fact that the total force (\ref{4.4})
vanishes in the system described by the static solution in Sec.~2,
substituting the relations (\ref{2.8}) and (\ref{2.9}) and taking
(\ref{4.1}) into account. 

We can also see the balance between the electric and the
scalar forces in
$(1+2)$ dimensions. In this case, we evaluate the forces at $f({\bf
x}^k)\sim 1$, i.e., $g_{ij}\sim\delta_{ij}$. 

The consideration here
is to show the physical implication; it should be recognized that the
static balance among the forces holds at any distance in the exact
solutions in any dimensions. 

Next let us consider the slow motion of
the point sources. Recently the slow motion of classical lumps or
solitons in many kind of field theoretical models has been
analyzed.\cite{11,12,13,14,15,16} The low-energy scattering of extreme
Reissner-Nordstrom black holes has been studied in Refs.~\cite{15} and
\cite{16}. It seems very useful to introduce the effective Lagrangian in
the low-energy limit in gravitating systems.\cite{14,15,16,17} We wish
to find the effective Lagrangian for the system which consists of two
maximally charged dilaton black holes in $(1+N)$ dimensions $(N\ge 3)$ .
The generalization to the many-body system would be a straightforword
task. 

Here the low-energy limit means the situation where any radiation
reaction can be ignored. Practically this means, in the effective
Lagrangian, keeping terms up to quadratic in the relative velocities of
the constituents of the system. 

We must be careful not to take double
counting of the contribution of different interactions. After some
straightforword calculation, we obtain an effective Lagrangian for two
point sources (where the velocity of light $c$ is explicitly indicated):
\begin{eqnarray}
L&=&\frac{1}{2}m_av_a^2\left(1+\frac{v_a^2}{4c^2}\right)
+\frac{1}{2}m_bv_b^2\left(1+\frac{v_b^2}{4c^2}\right)\nonumber  \\
&+&\frac{A_{N-1}}{4\pi(N-2)R_{ab}^{N-2}}\left[
Q_aQ_b-\sigma_a\sigma_b-\left(\frac{4\pi}{A_{N-1}}\right)^2
\frac{2(N-2)}{N-1}m_am_b\right]\nonumber \\
&+&\frac{A_{N-1}(v_a^2+v_b^2)}{8\pi c^2(N-2)R_{ab}^{N-2}}\left[
\left(\frac{4\pi}{A_{N-1}}\right)^2
\frac{2N}{N-1}m_am_b-\sigma_a\sigma_b\right]\nonumber \\
&+&\frac{A_{N-1}({\bf v}_a\cdot {\bf v}_b)}{8\pi
c^2(N-2)R_{ab}^{N-2}}\left[
Q_aQ_b+\sigma_a\sigma_b-\left(\frac{4\pi}{A_{N-1}}\right)^2
\frac{2(3N-2)}{N-1}m_am_b\right]\nonumber \\
&+&\frac{A_{N-1}({\bf n}\cdot {\bf v}_a)({\bf n}\cdot {\bf v}_b)}{8\pi
c^2 R_{ab}^{N-2}}\left[
Q_aQ_b-\sigma_a\sigma_b-\left(\frac{4\pi}{A_{N-1}}\right)^2
\frac{2(N-2)}{N-1}m_am_b\right]\nonumber \\
&+&O(1/R_{ab}^{N-1})\,,
\label{4.5}
\end{eqnarray}
where ${\bf n}$ is the unit vector in the direction $b-a$.
For the static configuration, we have found (for $i=a, b$)
\begin{equation}
m_i=\frac{A_{N-1}(N-1)}{8\pi(N-2+a^2)}\mu_i\,,
\label{4.6}
\end{equation}
\begin{equation}
|Q_i|=\left(
\frac{N-1}{2(N-2+a^2)}
\right)^{1/2}\mu_i\,,
\label{4.7}
\end{equation}
\begin{equation}
|\sigma_i|=a\left(
\frac{N-1}{2}
\right)^{1/2}\frac{1}{N-2+a^2}\mu_i\,.
\label{4.8}
\end{equation}

When (\ref{4.6}), (\ref{4.7}), and (\ref{4.8}) are substituted, the
second line of (4.5) vanishes trivially. This means no more than the
balance condition.

Further going on examining when each term in (\ref{4.5}) vanishes, we
encounter a simple result. The third and fourth lines of (\ref{4.5})
vanish when $a^2=N$. The fifth line vanishes identically, regardless of
the value of $a$. Thus all the interactions (of the order up to $v^2$)
disappear if and only if $N=a^2$. The case for $N=3$ can be related to
the Kaluza-Klein monopole\cite{14} through the duality. To analyze this
case closely is beyond the scope of the present paper and will be
reported in a separate publication.

Since the solution is a unique solution, point particles in general
Einstein-Maxwell dilaton systems feel the residual velocity-dependent
forces mutually.

\section{CONCLUSION}
We have found $(1+N)$-dimensional, static, multicentered solutions to
the Einstein-Maxwell equations coupled with dilaton. The solutions
stand for static configurations of many-body systems of maximally
charged dilaton black holes ($N\ge 3$). 

We have also found that there is no
interaction up to quadratic in $v/c$ between two extreme charged dilaton
black holes if and only if $N=a^2$. This case is related to the
Kaluza-Klein monopole ($N=3$) (Ref.~\cite{14}) via the duality in the
electromagnetism in four dimensions. 

We wish to report the thorough
investigation of the slow motion of the extreme dilaton black holes in
the future. We are also interested in the study of the connection
between the solutions obtained in this paper and supersymmetric models
in diverse dimensions.


\end{document}